\documentclass[acmsmall]{acmart}
\usepackage{listings}
\usepackage{mdframed}
\usepackage{graphicx}
\usepackage{multirow}
\usepackage{tabularx}
\usepackage{booktabs} 
\usepackage{paralist}
\usepackage{subcaption}
\usepackage{url}
\usepackage{amsmath}
\usepackage{amsfonts}
\usepackage{mathrsfs}
\usepackage[fixamsmath, disallowspaces]{mathtools} 
\usepackage{latexsym} 
\usepackage{textcomp} 
\usepackage{icomma} 
\usepackage{siunitx} 
\usepackage{caption}
\usepackage{float}
\restylefloat{table}
\usepackage{realboxes}
\usepackage{pgfplots}
\usepackage{subdepth}
\usepackage{circuitikz}
\usepackage{algpseudocode}
\usepackage{algorithm} 

\usepackage{longtable}
\usepackage{color}
\newcommand\newversion[1]{{\color{black} {#1}}}
\newcommand\newversiontwo[1]{{\color{black} {#1}}}
\usepackage{soul} 
\usepackage{tikz}
\usetikzlibrary{calc, arrows}
\pgfdeclarelayer{background}
\pgfdeclarelayer{foreground}
\pgfsetlayers{background,main,foreground}
\tikzset{
  every active component/.append style={
    draw=none,
    circuit symbol size=width 3 height 3,
  },
  >=stealth',
  LE marking/.style={
    rounded corners=2pt, line width=1pt, draw=white, rectangle, outer sep=0pt, inner sep=0pt, align=center,
  },
  scale text/.style={
    anchor=base, yshift=3pt,
  },
}

\graphicspath{{Figs/}}
\DeclareGraphicsExtensions{.pdf,.png,.jpg,.eps}
\pgfplotsset{compat=1.14}

\begin{document}
\title{RNNIDS: Enhancing Network Intrusion Detection Systems through Deep Learning\vspace{1cm}}
\noindent

\author{Soroush M. Sohi}
\affiliation{
  \institution{Security in Telecommunications \\
Technische Universit{\"a}t Berlin}
  \streetaddress{Ernst-Reuter-Platz 7}
  \city{Berlin}
  \state{Germany}
  \postcode{10587}}
\author{Jean-Pierre Seifert}
\affiliation{
  \institution{Security in Telecommunications \\
Technische Universit{\"a}t Berlin, and \\
Fraunhofer-Institut f{\"u}r Sichere Informationstechnologie}
  \streetaddress{Ernst-Reuter-Platz 7}
  \city{Berlin}
  \state{Germany}
  \postcode{10587}}
  
 \author{Fatemeh Ganji}
  \affiliation{\institution{Electrical and Computer Engineering Department\\Worcester Polytechnic Institute}
  \streetaddress{100 Institute Road}
  \city{Worcester}
  \state{USA}
  \postcode{01609-2280}}

\begin{abstract}
\begin{sloppypar}
Security of information passing through the Internet is threatened by today's most advanced malware ranging from orchestrated botnets to simpler polymorphic worms. 
These threats, as examples of zero-day attacks, are able to change their behavior several times in the early phases of their existence to bypass the network intrusion detection systems (NIDS).
In fact, even well-designed, and frequently-updated signature-based NIDS cannot detect the zero-day treats due to the lack of an adequate signature database, adaptive to intelligent attacks on the Internet. 
More importantly, having an NIDS, it should be tested on malicious traffic dataset that not only represents known attacks, but also can to some extent reflect the characteristics of unknown, zero-day attacks. 
Generating such traffic is identified in the literature as one of the main obstacles for evaluating the effectiveness of NIDS. 
To address these issues, we introduce RNNIDS that applies Recurrent Neural Networks (RNNs) to find complex patterns in attacks and \emph{generate} similar ones. 
In this regard, for the first time, we demonstrate that RNNs are helpful to generate new, unseen mutants of attacks as well as synthetic signatures from the most advanced malware to improve the intrusion detection rate. 
Besides, to further enhance the design of an NIDS, RNNs can be employed to generate malicious datasets containing, e.g., unseen mutants of a malware. 
To evaluate the feasibility of our approaches, we conduct extensive experiments by incorporating publicly available datasets, where we show a considerable improvement in the detection rate of an off-the-shelf NIDS (up to 16.67\%). 
\end{sloppypar}
\vspace{-0.2cm}
\end{abstract}

\keywords{\textbf{Network Security; Network Intrusion Detection Systems; Worm Mutants; Dataset Generation; Deep Learning; Recurrent Neural Networks.}\footnote{This is the author version of the paper accepted for publication in Computers \& Security, Elsevier (DOI: 10.1016\/j.cose.2020.102151).}} 
\maketitle
\section{Introduction}\label{sec:intro}
Nowadays we are witnessing rapidly escalating Internet threats, which have become increasingly mature as the Internet and its applications evolve. 
Today's Internet provides ubiquitous connectivity to a wide range of devices, with different operating systems, which indeed expands the available attack surface including several different attack vectors. 
As a prime example, according to the recent Symantec report\footnote{Symantec$^{\rm \text{\tiny TM}}$
has established the largest civilian threat collection network in the world, and has one of the most comprehensive collections of cyber security threat intelligence through the Symantec Global Intelligence Network composed of about 126 million attack sensors~\cite{symantec2017internet}.}, 
a significant increase can be observed in different classes of attacks, e.g., internet of things (IoT) devices (more than 600\%), new downloader variants (more than 92\%), etc.~\cite{symantec2018internet}. 
This increase has been partially fueled by the increased availability of user-friendly hacking tools, demanding solely superficial knowledge from attackers, as illustrated first by Lipson~\cite{lipson2002tracking} and further extended in~\cite{chester2020perspective}. 
Among malicious activities practiced by attackers, several classes of attacks can be recognized, for instance, Denial of Service (DoS)~\cite{loukas2010protection}, disclosure, manipulation, impersonation, and repudiation. 
These classes can be lumped together by an umbrella term, namely intrusion. 
Along with the emergence of increasingly sophisticated intrusions, intrusion detection systems (IDS) have been developed to cope with these threats.  
Regarding where or at which point an IDS is placed, two types of such systems can be distinguished: network intrusion detection systems (NIDS), and host intrusion detection systems (HIDS)~\cite{denning1987intrusion}. 
The latter is run on a device or an individual host in the network, whereas an NIDS is located within the network, at a strategic point, to monitor the traffic to and from all devices. 
Irrespective of this classification, the intrusion detection systems share some commonalities; first, they take advantage of the connectivity provided by networks, and secondly, they either use the known, specific patterns or apply anomaly detection techniques. 

Anomaly detection-based systems aim to establish behavior patterns, being different from the normal behavior of the system~\cite{wang2004anomalous}. 
Although these systems can be employed to detect previously unknown malicious activities, the main drawback of them is the high rate of false positives, i.e., a legitimate activity may be categorized as malicious~\cite{garcia2009anomaly}. 
On the other hand, signature-based detection systems attempt to match a known rule, a so-called \emph{signature}, with the contents of packets~\cite{denning1985requirements}. 
To this end, after receiving an incoming traffic, it undergoes a careful and continual process of analyzing and possibly generating signatures. 
This approach is similar to virus scanning mechanisms, where the database of the signatures, should be kept updated that can take from a few minutes to a few days. 
Despite this fact, the alternative method, i.e., anomaly-based detection, may require more processing power than signature-based ones. 
Additionally, although a less number of rules are required in comparison to signature-based systems, in highly dynamic environments it can be challenging to train anomaly-based detectors. 
Nevertheless, a careful signature definition can enable us to benefit from some of the advantages of using anomaly detector~\cite{mcgibney2004intrusion}. 
To this end, two major obstacles can be identified, as expalined below. 

\vspace{2pt}
\noindent\textbf{Generating signatures for intrusion variants:} The result of the signature generation process should be a set of \emph{effective} signatures, being narrowed down enough to characterize a specific attack, but flexible enough to detect some variations or modifications in an attack~\cite{kreibich2004honeycomb}. 
In other words, the probability of classifying benign traffic as malicious (i.e., false positives) must be low as well. 
Moreover, with high probability (i.e., low rate of false negatives) an effective signature should detect an attack, and possibly, its variants. 
The latter is related to the fact that although every day hundreds to thousands of signatures are generated and/ or updated by institutes and companies responsible for intrusion detection~\cite{buczak2015survey}, attackers create variants or \emph{mutants} of an intrusion to evade detection. 
Several different ways to generate an attack variant can be considered, ranging from simple interleaving of malicious data to sophisticated obfuscation of that. 
The point is that even a simple, small modification results in a completely new attack, similar to a new unknown zero-day attack, which requires performing the whole process of signature generation. 
In an ideal world, the signature generation should be conducted automatically so that an NIDS analyzes the incoming traffic and can distinguish between malicious and benign traffic with regard to inherent and unique characteristics of an attack, which offer a basis for generating the respective signatures. 
In addition to this, the problem with the derivation of an adequately specific signature are two of the greatest challenges faced by NIDS designers today. 

\vspace{3pt}
\noindent\textbf{Testing an NIDS by acquiring data as the ground truth:} Another serious obstacle to the implementation of an NIDS is how the effectiveness of the NIDS should be evaluated. 
For this purpose, traditionally a set of malicious data should be collected to serve as ground truth. 
Being close to the real traffic passing through real networks is one the specifications of such a set. 
DRAPA dataset~\cite{darpa1999mi}, KDD Cup~99~\cite{kdd99}, and CDX~\cite{cdx09,sangster2009toward} can be considered as attempts to address this issue. 
However, they suffer from a lack of nearly real traffic data, and they do not include traffic for all of the network protocols and all variants of attacks in today's world~\cite{zuech2015new}. 
As an example of remedies for this, in~\cite{song2011statistical} another dataset called ``Kyoto~2006+'' for research purposes has been offered. 
In one of the most recent attempts, the authors of \cite{beigi2014towards} have combined several types of malicious datasets available on the Internet to produce a mixed dataset, which satisfies the requirement of having a wide range of various attacks. 
They have applied a method called ``overlay methodology'' to produce synthetic dataset~\cite{aviv2011challenges}.  
According to this method, data related to malware activities is merged with benign data by sending the data to the machines on an external network. 

Despite the above, according to at least two main reasons, collecting a set of \emph{malicious data} providing the ground truth is not straightforward~\cite{aviv2011challenges}. 
First and foremost, typical network traces contain sensitive information, which is carefully controlled and cannot be shared with other parties. 
This is due to the fact that even after a cautious anonymization process, it can be still possible to extract some sensitive information from the data~\cite{narayanan2008robust}. 
Second, given that an NIDS is tested by feeding a small set of data, due to the heterogeneous nature of the real-world network traces, the results of the test cannot be representative. 
Hence, it has been proposed to use synthetic data traces, see, e.g.,~\cite{vishwanath2009swing}. 
Nevertheless, the impact of possible biases and limitation with respect to realism should be considered in this case~\cite{aviv2011challenges}. 
More importantly, the synthetically generated traces mainly follow known trends such as distributions of users, applications as well as the network behavior, and do not reflect further detailed characteristics of the traffic, e.g., the payload that is crucial for efficient intrusion detection. 
Therefore, it is of great importance to establish a methodology which can be used to achieve realism to a higher degree. 

Furthermore, and more crucially, when a detector is implemented and its effectiveness against zero-day attacks should be tested, it is necessary to generate a traffic pool reflecting the nature of the zero-day attacks. 
In the literature, this has not been completely addressed so far. 
As a prime example,~\cite{holm2014signature} re-defines zero-day attacks as the attacks discovered after the NIDS under study is released. 
By taking this into account, it has been shown that a signature-based NIDS, namely SNORT~\cite{snort1998}, can detect a set of zero-day attacks. 
Unfortunately, the issue with ground truth remains a critical challenge. 
\begin{figure}[t]
\begin{center}
\includegraphics [width=0.9\columnwidth]{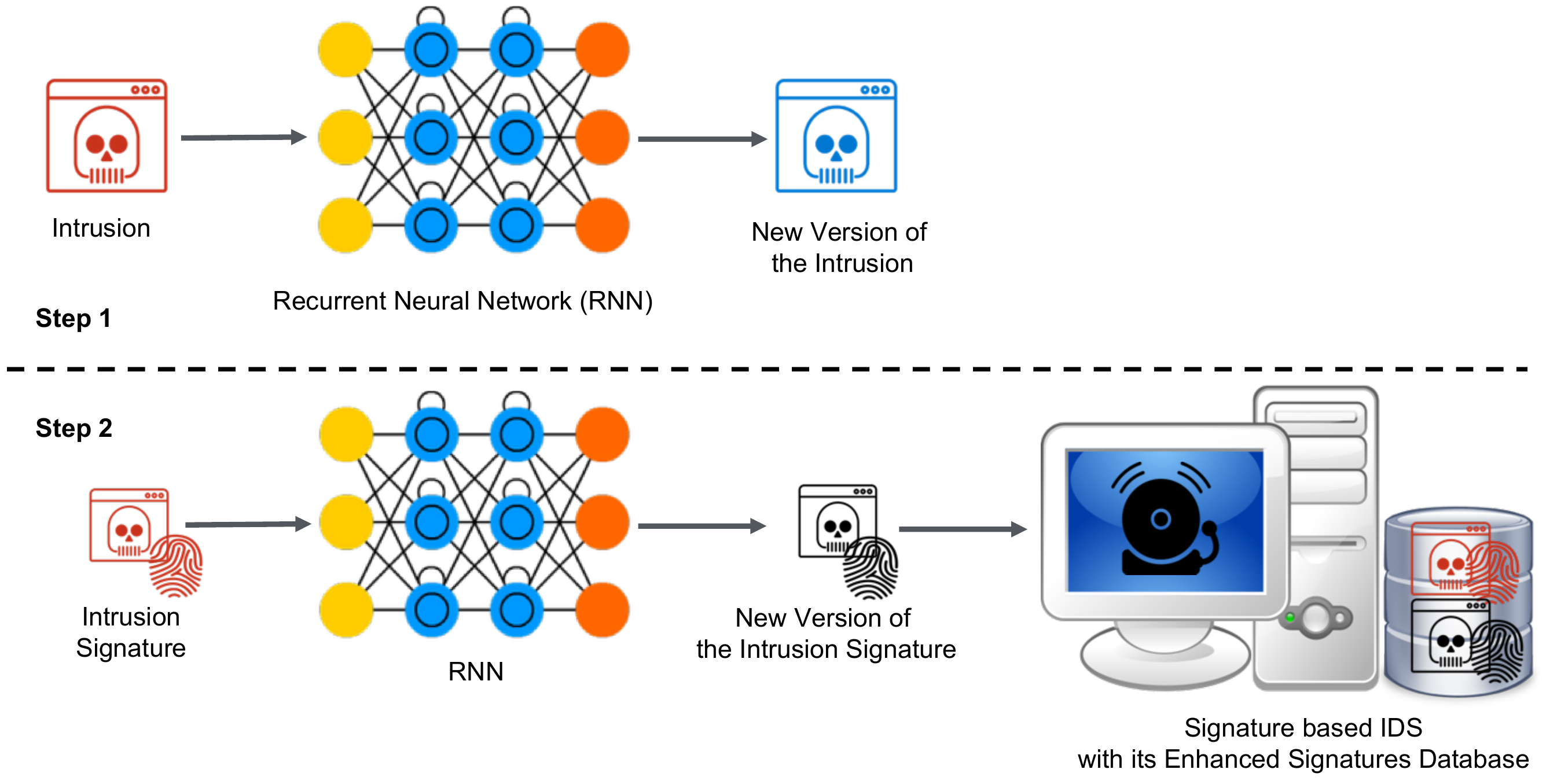}
\caption{Two steps that are taken in our framework to present how deep learning, namely, Recurrent Neural Networks (RNNs),  can improve the effectiveness of an NIDS.
The first step shown in this figure corresponds to our first contribution, namely, generating attack mutants. 
The second step is taken by us to generate not only attack signatures, but also synthetic data that can be used to evaluate the performance of an existing NIDS. \vspace{-5pt}} 
\label{fig:framework}
\end{center}
\end{figure}

\noindent\textbf{Our contributions:} This work aims at removing obstacles mentioned above by contributing to the following points. 

\vspace{2pt}
\noindent\textbf{(1) Generating malware mutants: } 
By applying a novel and sound method, we demonstrate how mutants of a malware can be generated. 
This has been illustrated in Figure~\ref{fig:framework} and marked as Step~1. 
Our method relies on the fact that an unknown pattern, so-called grammar, can be extracted and learned by a recurrent neural network (RNN). 
This fact enables us to generate new, unseen sequences, i.e., attack mutants. 
These new variants of a malware can be indeed helpful, when there is a need to generate a dataset that serves as the ground truth. 

\vspace{2pt}
\noindent\textbf{(2) Generating attack signatures: }
As another example of how RNNs can be used in intrusion detection, we demonstrate that by extracting the signatures of an NIDS and feeding them into an RNN it is possible to generate synthetic signatures. 
These signatures can be further added to the database of an NIDS to improve its performance in terms of finding unseen mutants of an attack, as shown in Figure~\ref{fig:framework} (see Step~2\newversion{, for the results, see Section~\ref{sec:res_bro} ``Experiment~1'')}. 

\vspace{2pt}
\noindent\textbf{(3) Synthetic data generation: }
The methodology that we apply in this paper can represent a change of direction for generating synthetic data. 
More specifically, we show that along with applying the overlay methodology by using synthetic signatures and mutants generated by an RNN, it is possible to generate synthetic data to test an NIDS.
We stress that when evaluating the performance of the NIDS by feeding this synthetic data, the synthetic signatures used to generate that should have been removed from the database of the NIDS (see Section~\ref{sec:res_bro} for more details). 

Last but not least, we verify the relevance of the above claims empirically by applying the proposed concept to one of the well-studied, off-the-shelf NIDS, namely Zeek \newversiontwo{(formerly Bro, henceforth called Bro)} NIDS~\cite{zeek2020}. 
According to our experimental results, when incorporating a publicly accessible dataset, the performance of Bro can be improved by up to about 17\%. 
Nevertheless, the scope of our work is not restricted to this NIDS.  
In other words, our methodology is applicable in other scenarios, where a signature-based NIDS is employed to find an intrusion. 

\noindent\textbf{Organization: } 
This paper is structured as follows.
Section~\ref{sec:prelim} introduces the notations and concept used to describe our scheme. 
Section~\ref{sec:backgroundonRNN} provides a brief overview of the mechanism of Recurrent Neural Networks (RNNs) and how they can be used to generate texts.  
In Section~\ref{sec:approach}, we elaborate on our methodology applied to generate new mutants of malware as shown in Step~1 in Figure~\ref{fig:framework} (see Section~\ref{sec:approach_worm}), while Step~2 in our framework is described in Section~\ref{sec:approach_sig}.  
Moreover, Section~\ref{sec:exp_desc} is devoted to the definition of the metrics and the description of the experimental setup, and the design of the experiments. 
We present our experimental results in Section~\ref{sec:results}, and finally, conclude the paper in Section~\ref{sec:conclusion}. 
 
\section{Notations and Definitions}\label{sec:prelim}
Although we assume that the reader can be familiar with the concept of regular languages, Deterministic Finite Automata (DFA), and deep learning, we define the functions and notations used throughout this paper. 
Note that standard notation is used here, as found in \cite{Goodfellow-et-al-2016}.

\vspace{2pt}
\noindent\textbf{Preliminaries on Formal Languages and DFAs}

Consider the alphabet $\Sigma=\{0,1\}$ and the set of all strings $\Sigma^*$ over $\Sigma$.
A set $L \subseteq \Sigma^*$ is called a language over the alphabet $\Sigma$. 

A grammar is a $4$-tuple $(N, V, P, S)$, with the sets of non-terminal ($N$) and terminal ($V$) vocabularies, i.e., strings, a finite set of production rules ($P$), and the start symbol ($S$). 
For each grammar, there exists a language corresponding to that as well as an automaton recognizing the strings of that grammar. 
Our focus is on deterministic and regular grammars, which can be recognized by a DFA. 

A DFA $A$ is defined by 
$A=(Q, \delta, \Sigma, q_0, F)$ over the alphabet $\Sigma$, where $Q$ is the set of states, the initial state is denoted by $q_0$, and the accepting states are $F \subseteq Q$.
$\delta : Q \times \Sigma \to Q$ is the transition function defined as follows.
For all $q \in Q$, $a \in \Sigma$ and $c\in \Sigma^*$, we have 
$\delta(q, \lambda) = q$ and its canonical continuation to $\Sigma^*$, i.e., $\delta(q, ac)=\delta(\delta(q, a), c)$. 
Giving strings to $A$, it accepts a set of them, called its accepted language $\textbf{L}(A) := \{c \in \Sigma^* \mid \delta(q_0, c)\in F\}$, 
i.e., a regular language. 

From another perspective, particularly by taking a bottom-up approach, regular languages can be described by their respective regular expressions (often called, regex). 
In an informal and intuitive way, regular expressions enable us to begin with building blocks and combine them to generate other regular expressions. 
These building blocks are regular expressions representing the empty language $\lbrace\emptyset \rbrace$, the language of the empty string $\lbrace \lambda \rbrace$ with $|\lambda|=0$, and the languages of the sets containing only one alphabet of $\Sigma$, i.e., $\lbrace a\rbrace$ s.t. $a \in\Sigma $. 
In order to combine these building blocks and obtain new regular expression, we can apply the concatenation, the union, the Kleene closure, and the intersection operators (for more details see, e.g.,~\cite{hopcroft2008introduction}). 
It has been proved that $L$ is a regular language iff there is a regular expression $R$ such that $\textbf{L}(R) = L$~\cite{hopcroft2008introduction}.
Moreover, it is known that DFAs and regular expressions are equivalent, i.e., if a language $L$ is built up by regular expressions s.t. $\textbf{L}(R) = L$, there exists a DFA $A$ that accepts this language, i.e., $L=\textbf{L}(A)$~\cite{hopcroft2008introduction}. 

\noindent\textbf{Notations Related to Deep Neural Networks:} 

One of the functions widely applied when working with deep learning models is (logistic) sigmoid:  
\mbox{$\sigma (x) = 1/ 1+ \exp (-x)$}.
In addition to the sigmoid function, the hyperbolic tangent function, $\tanh(\cdot)$, is used as an activation function in deep networks. 
\begin{figure}[t]
\begin{center}
\includegraphics [width=0.5\columnwidth]{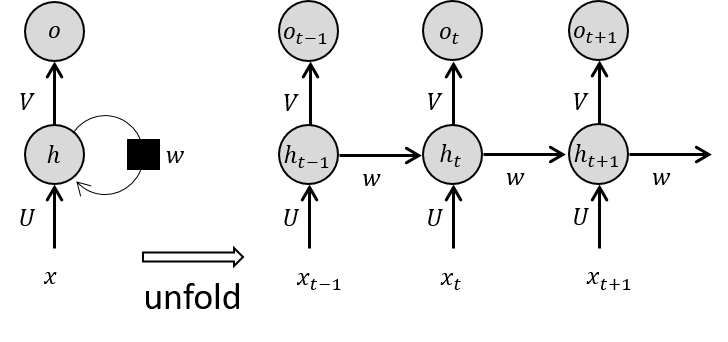}
\caption{In an RNN, an input from the previous state of the hidden layer is fed into the next hidden layer designated with a weight $w$. Weights corresponding to the input and the output layers are $U$ and $V$~\cite{lecun2015deep}.}
\label{fig:unfoldedRNN}
\end{center}
\end{figure}

\begin{figure*}[t]
\begin{center}
    \begin{subfigure}[b]{0.31\textwidth}
    \begin{center}
        \includegraphics[width=\textwidth]{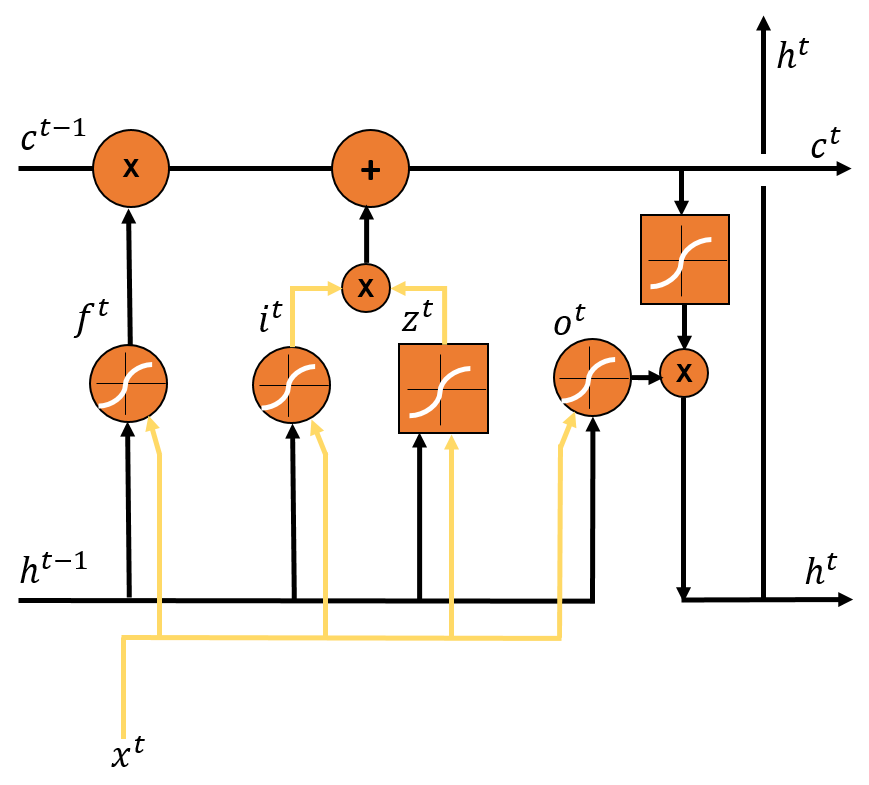}
        \caption{ }
        \label{fig:LSTMCell}
        \end{center}
    \end{subfigure}
    \hspace*{0.75cm}
    \begin{subfigure}[b]{0.62\textwidth}
    \begin{center}
\includegraphics [width=\textwidth]{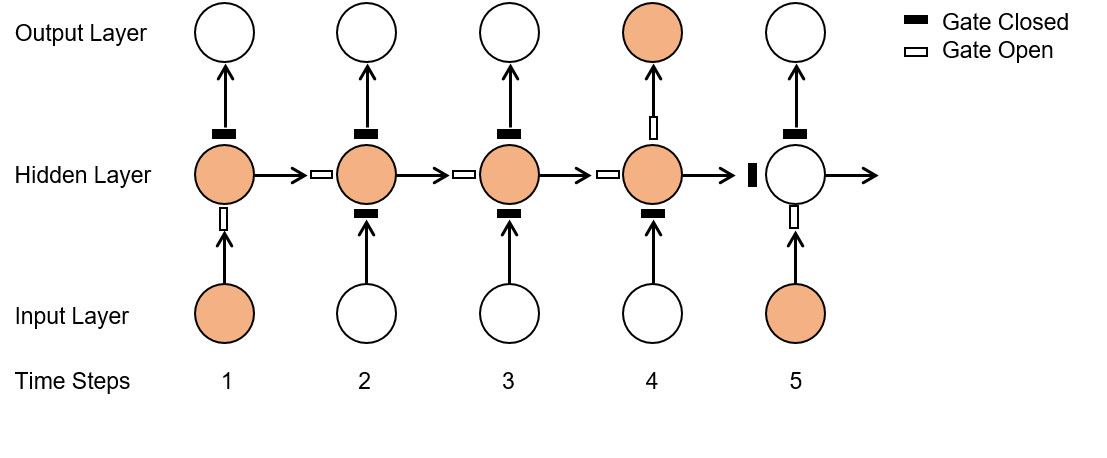}
\caption{ }
\label{fig:LSTMGates}
\end{center}
\end{subfigure}
\caption{(a) Structure of an LSTM memory cell. The previous states of the cell and the hidden states could be erased, updated, and read by the gates. Squares are hyperbolic tangent activation functions and circles are logistic sigmoid activation functions, (b) Input, output, and forget gates in an RNN with LSTM cells. 
These gates control the flow of information in different time steps~\cite{gers1999learning,graves2012supervised}.}
\end{center}
\end{figure*}
\section{Brief Overview of Recurrent Neural Networks (RNNs)} \label{sec:backgroundonRNN}
Although the concept and applications of recurrent neural networks (RNNs) are nowadays part of the common knowledge in our community, as a self-contained paper, our work briefly introduce the most important concepts required to understand our method. 

The concept of deep neural networks, as an approach to artificial intelligence (AI), has been first motivated by problems that could not be addressed properly in traditional machine learning. 
As examples, speech recognition and generating text~\cite {sutskever2011generating} can be mentioned~\cite{Goodfellow-et-al-2016}. 
One of the most prominent, and widely-applied neural networks is feed-forward networks, where via a series of weights the inputs is given to the outputs directly. 
Although being a very powerful tool for several applications, feed-forward networks cannot cope with issues, in which a sequence of inputs should be processed. 
To address this, RNNs have been proposed (see, e.g.,~\cite{schmidhuber1993netzwerkarchitekturen}), which exhibit temporal characteristic fulfilling the conditions for processing sequences of data. 

In addition to being superior in learning sequences~\cite{graves2012supervised}, RNNs can be employed to generate sequences that are similar to training sequence~\cite{lecun2015deep}. 
In Comparison to feed-forward neural networks, each stage of an RNN has an input coming from the previous state, in a similar way to memorizing the history of all the past states (see Figure~\ref{fig:unfoldedRNN}). 
In the fields of speech recognition and Natural Language Processing (NLP), RNNs are usually used to predict the next word in a sequence. 

A major problem with RNNs is the vanishing or exploding gradients in back-propagation process~\cite{Goodfellow-et-al-2016}. 
One of the first designed RNN (so-called vanilla RNN), which is the simplest one suffers from this problem. 
In other words, legacy, vanilla version of RNNs could not save the memory for a long time and it could vanish during the time. 
Consider an RNN that is trained over 10000 samples and builds up the memory upon those. 
As each new sample with new attributes arrives in a sequence, the memory is overwritten after a while and it forgets the previous memory. 
Long Short-Term Memory (LSTM) RNNs have attempted to solve this problem~\cite{graves2012supervised}.  
The concept of LSTM was first introduced in 1997 by Hochreiter and Schmidhuber~\cite{hochreiter1997long}. 
An LSTM RNN can be turned to a gated RNNs, which has the capability to learn for a long time, and, when needed, it has the gates to forget the memory and learn again based on new inputs~\cite{gers1999learning}. 
These gates also allow RNNs to pass the information unchanged to other layers in the network, i.e., featuring read, write and clear functions as illustrated in Figure~\ref{fig:LSTMGates}. 
When a gate is open, it allows the information propagation through the gate. 
More formally, in Figure~\ref{fig:LSTMCell}, $i^t$ , $f^t$, and $o^t$ are the output signals of the corresponding input, forget, and output layers at the time step $t$, whereas $h^t$ is an updated hidden state value used for the next time step and the output to the upper layer at that time step. 
Moreover, $c^t$ is the cell state (so-called, memory state) fed into the next time step, and similarly, $c^{t-1}$ is the cell state coming from the previous time step.
And let $W_f$, $W_i$, $W_z$, and $W_o$ be the input weights related to each gates, and in a similar fashion, $R_f$, $R_i$, $R_z$, and $R_o$ be corresponding to the recurrent inputs from the previous states.
Now the output of the input layer at the time step $t$ can be formulated as follows. 
$$i^t=\sigma (W_ix^t+ R_ih^{t-1}+b_i),$$
where $b_i$ is the bias for the input gate. 
By substituting the weights and biases for the forget and output gates in the above equation, we obtain the outputs of the other layers, namely $f^t$, $o^t$, and $z^t$. 
Finally, the cell state at the time step $t$ is $c^t=f^t\cdot c^{t-1}+i^t \cdot z^t$, and the updated hidden state $h^t=o^t\cdot \tanh (c^t)$. 

The structure explained above enables LSTM RNNs to cope with long-term dependencies more effectively than the simple recurrent networks~\cite{Goodfellow-et-al-2016}. 
Therefore, LSTM RNNs have become powerful and widely accepted tools for speech~\cite{graves2013speech} and handwriting recognition as well as text generation~\cite{graves2012supervised}. 

\subsection{Application of RNNs in Text Generation}
Sequence learning offers a wide range of applications, for instance, natural language processing, time series prediction, and DNA sequencing. 
Due to these applications, sequence learning can be considered as a discipline or of a particular line of research developed to address problems such as sequence prediction, sequence recognition, etc.~\cite{sun2001sequence}. 
Among those problems, sequence prediction (mostly referred to as sequence generation) attracts a great deal of attention because of its applications in several domains of study, e.g., human-machine interaction~\cite{laird1994discrete}. 

After the development of RNNs in 1980s, they are widely used in enormous studies related to sequence prediction. 
Giles et al. have pursued this line of research by demonstrating fundamental strength of RNNs and their close relationship to deterministic finite automata, see, e.g.,~\cite{giles1990higher,giles1992extracting,giles1992learning,omlin1996constructing}. 
They began with a formal model of sequences, i.e., formal grammar and machines generating and recognizing them, namely their respective automata. 
More specifically, they have proved that RNNs can be trained to simulate deterministic finite automata (DFA) and recognize their corresponding grammar~\cite{giles1990higher}. 
Moreover, it has been demonstrated that, even unknown, (small) grammars can be learned and further extracted from the RNN (i.e., generating the corresponding DFA)~\cite{giles1992extracting}. 
It is worth noting here that the problem of grammatical inference is proved to be NP-complete~\cite{gold1978complexity}, and results presented by Giles at al. are in line with heuristic approaches attempting to address grammatical inference (see, e.g.,~\cite{angluin1983inductive}). 

\noindent\textbf{Comparison to Generative Adversarial Networks (GANs): }
One may argue why GANs have not been taken into account in our framework. 
The point is that GANs have achieved a promising result in the image generation domain; however, for text generation, several attempts have failed as discussed below. 
When combined with RNNs, GANs cannot be trained effectively due to the non-differentiable nature of generating discrete symbols. 
Hence, pre-training and joint training methods have been suggested to tackle this issue, see~\cite{press2017language} for an exhaustive discussion.
On the other hand, CNN-based GANs are also studied for text generation, see, e.g.,~\cite{gulrajani2017improved}, which lead to less desirable results, namely the generated text contains spelling errors and has had little coherence. 
Therefore, for our purpose, we stick to RNNs, which not only require relatively less effort to be trained, but also generate more coherent text.

\section{Methodology: Applications of RNNs in Intrusion Detection}\label{sec:approach}
This section covers the methodology applied to generate mutants of worms as well as synthetic signatures used in an NIDS. 
The main idea behind these observations is to demonstrate not only the possibility of generating new variants of worms and signatures from a small number of examples (see Figure~\ref{fig:framework}, Step~1), but also to show how these variants can extend the dataset of an NIDS to improve its detection rate (Step~2 in Figure~\ref{fig:framework}). 
More specifically, this section is mainly devoted to theoretical and practical approaches that we apply to generate those variants. 
In fact, we show that an LSTM RNN can extract the ``deep structural properties'' of an encoded worm and a signature to generate variants of them.  
\subsection{Generating New Mutants of Polymorphic Worm}\label{sec:approach_worm}
As discussed in Section~\ref{sec:intro}, one of the major obstacles that signature-based NIDS encounter is the lack of knowledge about new variants of an attack. 
As an example of attacks with enormous variants, we focus on polymorphic worms. 
It is known that these worms are able to change their behavior through generating mutants or variants of themselves to pass through NIDS without being detected. 
They attempt to hide their encrypted malicious code in all variants of themselves. 
The notion of ``generating mutants'' of an exploit has been introduced in the literature, which should not be confused with mutation testing. 
Following the reasoning provided in~\cite{vigna2004testing}, in our scenario we generate mutants of an attack and the targeted NIDS remains untouched, in contrast to mutation testing approaches. 
An example of a mutant generator for NIDS, compatible with our scenario, has been introduced in~\cite{vigna2004testing}. 
Similar to our goal, ~\cite{vigna2004testing} aims at evaluating the effectiveness of an NIDS by feeding mutants of a known attack. 
Nonetheless, the main drawback of their proposed method is that the mutation mechanism should be maintained continuously. 
In other words, the existing mechanisms have to be frequently updated, and more importantly, the parameters reflecting the nature of an attack should be extracted before generation process begins. 
This section of our paper attempts to address these issues. 

\begin{figure}[t]
\begin{center}
\includegraphics [width=0.8\textwidth]{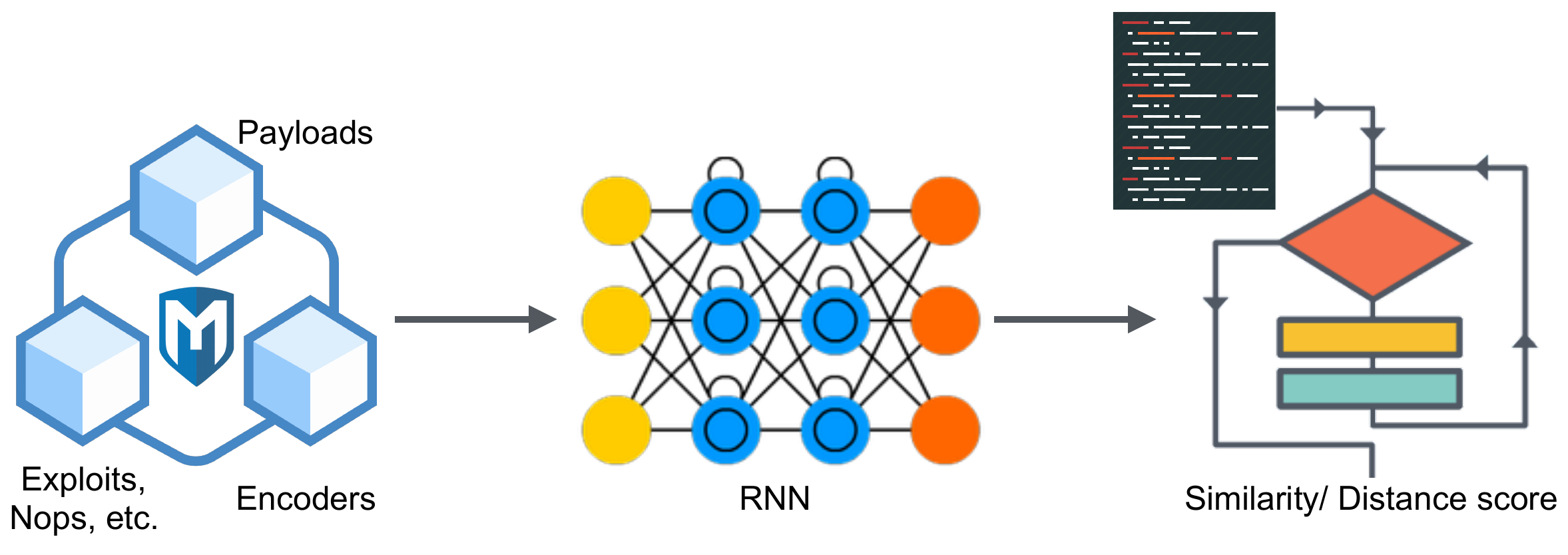}
\caption{Steps taken in our approach: (from left to right) generating the worms and their mutants using the Metasploit framework, then feeding them into an RNN, and finally compare the results in terms of the similarity.} 
\label{fig:app_worms}
\end{center}
\end{figure}

More specifically, the primary goal of our approach described in this section is to observe if new, unseen variants or mutants of a known worm can be generated by an LSTM\footnote{Hereafter, we may use the terms ``RNN'' and ``LSTM'' interchangeably, since we solely focus on and implement LSTM RNNs.}, given some previously collected variants of that. 
In particular, we take the steps illustrated in Figure~\ref{fig:app_worms}. 
\newversion{To provide proof of concept of how our approach can be applied, here we focus on different state-of-the-art encoder engines implemented in ADMMutate~\cite{adm1990k2} as well as the Metasploit framework ~\cite{metasploit}
, namely the XOR encoder~\cite{metasploit_xor} and the Shikata Ga Nai (in Japanese means ``nothing can be done about it'') encoder ~\cite{metasploit_shikata}
. }
Using such encoders are crucial since when an exploit in the form of, e.g., a shellcode is generated, it cannot be directly used. 
But it should be encoded to obtain a pure alphanumeric code, by removing bad characters (e.g., null bytes). 
Moreover, the encoded exploit may suit 64-bit target systems, which can be achieved by deploying an encoder. 
For instance, in the Metasploit framework, the XOR encoder ~\cite{metasploit_xor} 
employs an 8-byte key and leverages relative addressing used by x64 operating systems, whereas the Shikata Ga Nai encoder is a polymorphic XOR additive feedback encoder for the x86 architecture. 
The decoder stub of this encoder is generated by substituting the instructions and ordering the blocks dynamically. 
In this regard, after each iteration different outputs are generated in the hope that signature recognition can be prevented.
Furthermore, the key used by the encoder is modified through additive feedback.
Additionally, the decoder stub is also obfuscated (for more details, see ~\cite{farley2014codext} 
). 
Beside the XOR and Shikata Ga Nai encoders, we also consider the ADMMutate engine exhibiting the following features. 
The payloads are encoded by applying 16-bit sliding keys. 
Moreover, the ADMMutate supports randomized NOP generation, banned characters, insertion of non-destructive junk instructions and the reordering/substitution of code as well as polymorphic payload decoder generation with multiple code paths~\cite{vigna2004testing}.

Comparing the Shikata Ga Nai encoder with the XOR encoder, as indicated by Metasploit ~\cite{metasploit}
, it is expected that Shikata Ga Nai encoder outperforms the XOR encoder. 
However, successful detection of exploits using these encoders have been reported in the literature, see, e.g., ~\cite{song2007infeasibility} 
. 
This is due to the fact that these encoders leave traces in the encoded payload that can be used to detect the exploit. 
A visual, effective procedure proposed in ~\cite{song2007infeasibility} 
to illustrate such traces for, e.g., Shikata Ga Nai encoder. 
We follow that procedure to see if for the ADMMutate, the XOR, and the Shikata Ga Nai encoders any pattern can be observed, which can be further used by an LSTM to generate new mutant of a worm. 

\begin{figure}[t]
\begin{center}
\includegraphics [width=0.9\textwidth]{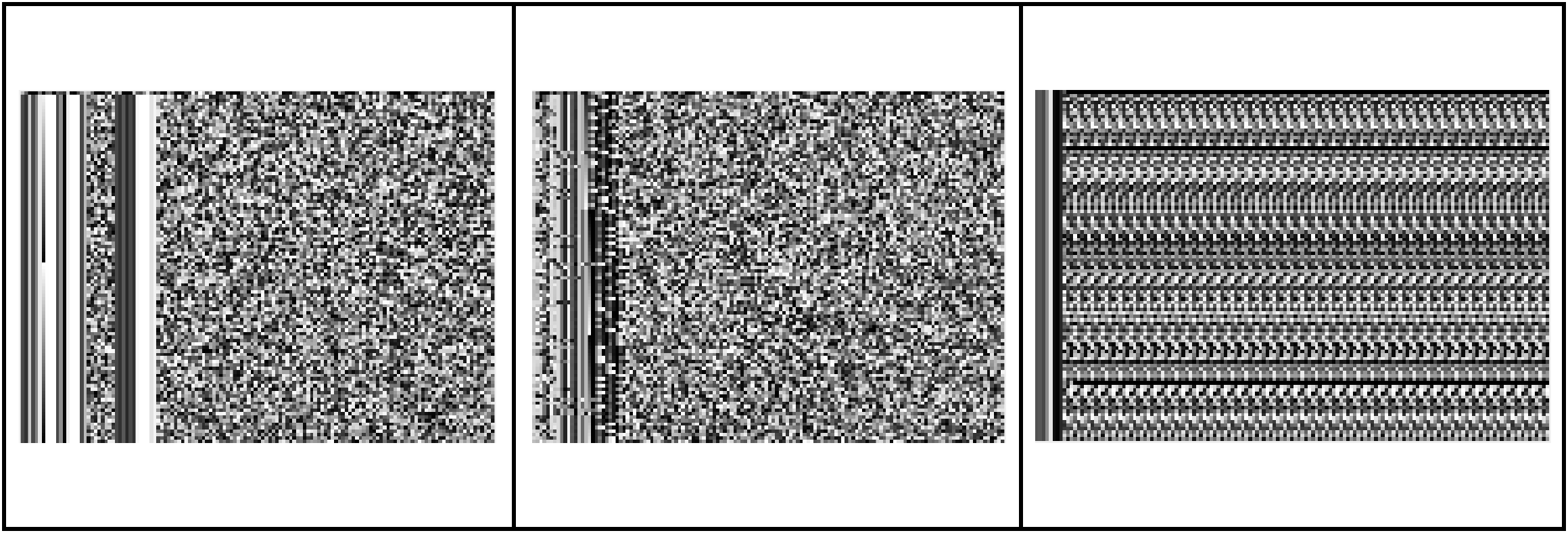}
\caption{So-called spectral images illustrating the patterns (the visible columns) formed by the bytes
repeated in the same places within all encoded bash files. 
In all examples, each row of the pixels shows a decoder generated by an engine, whereas each pixel corresponds to a byte of that decoder.
Although in comparison to the XOR encoder (left) such patterns are less visible for the Shikata Ga Nai (middle) encoder and ADMMutate (right), they can be still distinguished.} 
\label{fig:allworms}
\end{center}
\end{figure}
To this end, we define the following setting.
\newversion{To generate \emph{examples} of} exploits in the form of bash files, we apply the Shikata Ga Nai, XOR, and ADMMutate encoders to the OS X x64 Shell Bind TCP payload~\cite{metasploit_tcp_payload}
, which binds an arbitrary command to a port chosen by the attacker. 
Moreover, the number of iterations for each encoder varies from zero (i.e., without encoding) to 100 so that 101 bash scripts are generated. 
These bash files are sorted in matrices in a row-wise manner. 
Note that we add padding (all zero values) at the end of the rows to obtain rows with the same length since for these encoders, the outputs of these encoders are non-fixed length bash files. 
Afterwards, the matrices are displayed as grayscale images, where a byte value of 0x00 (0xFF) corresponds to a black (white) pixel. 
Figure~\ref{fig:allworms} shows portions of the worms encoded by the Shikata Ga Nai, XOR, and ADMMutate encoders. 

The most important message to be conveyed here is that the bash files generated by applying the above-mentioned encoding engines do not comprise of completely, and truly random values. 
In other words, there are some patterns, i.e., the bytes repeated in the same places within all bash files, which form the visible columns in Figure~\ref{fig:allworms}. 
\newversion{We stress that while the results discussed above should not be extrapolated to all payloads/encoders, they show patterns that can be useful for generating new variants of an attack. }
This is in line with what has been observed in~\cite{li2006hamsa}, where the approach relying on finding unavoidable byte patterns is called ``content-based''. 
According to this definition, our approach can be classified as content-based as well. 
The main advantage of such approaches is that there exists no dependency on protocol or server information~\cite{li2006hamsa}. 
As well described by Li et al., in the real-world, the protocol frame part and the control data of the worm cannot be manipulated by an attacker to obtain a new mutant of a worm~\cite{li2006hamsa}. 
These parts create patterns that can be indeed used by an LSTM network to generate similar and unseen bash files as mutants of the original payload. 

More formally, by providing examples of worms we train our LSTMs to learn and extract rules corresponding to a DFA $A$, as proved by Giles et al.~\cite{giles1992extracting}. 
This can be explained by the fact that the grammar of bounded-length strings is regular. 
Recall that there is a one-to-one mapping between a DFA and its grammar. 
Let the grammar corresponding to the DFA $A$ be denoted by $(N, V, P, S)$, and the language generated by this grammar be $L'$. 
In fact, $L'=Enc(L)$, where $Enc(\cdot)$ represent the encoding operation applied to encode the original language generating a worm $L$. 
After training, the grammar $(N, V, P, S)$ is used by the LSTM to generate unseen strings, which belong to $L'$. 

Up to this point, we have discussed the first two steps of our approach shown in Figure~\ref{fig:app_worms}, namely generating the original worms and their respective mutants by applying the concept of LSTM. 
In Section~\ref{sec:results}, we discuss the last step that is computing the similarity or the distance score. 
Additionally, we provide the results that we obtain through empirical evaluation of our approach in Section~\ref{sec:results}. 

\subsection{Generating Synthetic Signatures}\label{sec:approach_sig}
\begin{figure}[t]
\begin{center}
\includegraphics [width=0.8\textwidth]{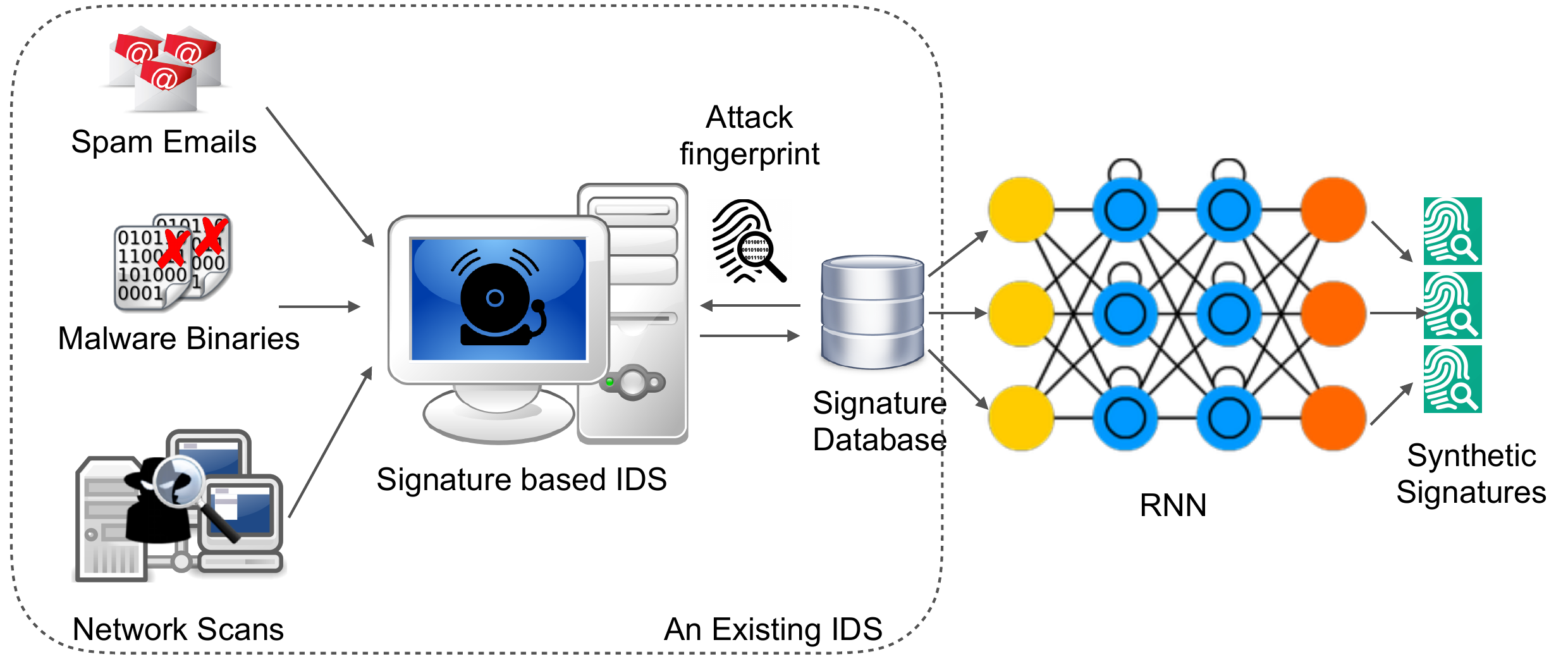}
\caption{Our framework to extend and enhance an existing NIDS: (from left to right) extraction of the signatures stored in the database of the NIDS, then feeding them into an RNN, and at the end the synthetic signatures generated by the RNN used to extend the database of the signatures.} 
\label{fig:app_sig}
\end{center}
\end{figure}

As discussed in Section~\ref{sec:intro}, we mainly focus on signature-based intrusion detection methods. 
\newversion{Although both signature- and anomaly-based classes of NIDS may suffer from false positives, signature-based systems can exhibit a lower rate of that~\cite{mcgibney2004intrusion,buczak2015survey,andress2014basics}. 
The popularity of signature-based NIDS could stem from this fact.}
. 
As a prime example of such popular NIDS, one can mention Zeek (previously called Bro)~\cite{zeek2020} 
and Snort~\cite{snort1998}. 
Albeit significant developments and improvements in the design of signature-based NIDS, attackers are often steps ahead of these defense systems. 
This can be explained by the fact that nowadays, a wide range of methods and tools are accessible by an attacker, and the attack patterns and behaviors change rapidly.

Generation of signatures for known attacks has become a mature area of research. 
In this context, numerous approaches applying various methodologies, ranging from machine learning and data mining (see, for example,~\cite{lee1998data,tsai2009intrusion}) to matrix factorization~\cite{krueger2010asap}, have been proposed in the literature. 
Nevertheless, our work does not share a great deal of commonality with these approaches. 
It is due to our approach that can be seen as an add-on to each and every NIDS to improve the effectiveness of that. 
One can observe a closer relationship between our approach and studies discussing synthetic data generation for evaluation of NIDS, e.g.,~\cite{bar2016scalable}. 
A comparison between the method employed in~\cite{bar2016scalable} and our approach can explain why we shift our focus to RNNs from Markov chain models, as proposed by Bar et al.~\cite{bar2016scalable}. 
Although being a predominant framework for different applications, Markov chain models suffer, in particular, from the following disadvantages. 
They need a significant amount of knowledge of a task performed by them. 
Moreover, the assumptions underlying the design of these chains can be questionable, e.g., the dependency assumptions. 
Our RNN-based approach attempts to address these issue. 

Although we do not limit our approach to Bro NIDS, we take that as a prime example of a popular signature-based NIDS.
Bro is an open source application running on Unix based systems. 
It has its own scripting language, enabling users to write the user-specific events and logs scripts. 
Additionally, Bro has the capability of changing the structure at many levels according to the user demands. 
It supports many popular transmission protocols and is able to analyze them. 
Moreover, it can extract the data packets and match the data with the predefined rules. 
Last but not least, Bro supports regular expression for pattern matching ~\cite{bro2018introduction}
. 
In more details, in each signature embedded in Bro, the information related to the payload of the attack is represented by its corresponding regular expression.

The latter feature of Bro is interesting from the point of view of our methodology.  
As can be seen in Figure~\ref{fig:app_sig}, which illustrates our roadmap for generating synthetic signatures, the signatures of Bro is fed into an LSTM. 
More precisely, the regular expression $R$ related to the payload of an attack is given to the LSTM. 
When the LSTM learns the DFA associated with $R$, it extract the DFA $A$ such that $\textbf{L}(R) = L$, and equivalently, $L=\textbf{L}(A)$. 
In the next step, the LSTM uses the information about $L$, and $A$ to generate another valid regular expression $R'$, where 
$\textbf{L}(R') = L.$
Afterwards, the languages $L$ and $L'=\textbf{L}(R')$ can be combined by using, e.g., the union operator to obtain an enhanced regular language $L \cup L'$, with the associated regular expression $R \cup R'$. 
This union can be used as a substitute for $R$ to improve the detection rate of Bro. 
We evaluate this quantitatively and present the results of deploying this method in Section~\ref{sec:results}.

\section{Experiment Description}\label{sec:exp_desc}
Before providing the results achieved by experimentally evaluating the performance of our methods (see Section~\ref{sec:approach}), here we first introduce the metrics, experimental setup, and explain the design of the experiment. 
\subsection{Metrics}\label{sec:res_metrics}
We define two comprehensive sets of metrics used for (1) computing the distance, or equivalently, comparing the similarity between the original sample (i.e., a set of original worms or Bro signatures) and samples generated by an RNN, and (2) evaluating the performance of the enhanced Bro in practice. 

\vspace{2pt}
\noindent\textbf{Similarity Comparison (Distance Measuring)}

In order to examine how good the outputs of the RNN are, i.e., how far they are from the original samples, two famous similarity metrics are taken into account. 

\emph{Levenshtein distance}: this distance metric defines how many substitutions, deletions or insertions are required to transform one string to another one. 
To assess the similarity of two strings, this metric has been used in the intrusion detection-related literature, e.g.,~\cite{tang2009using,cesare2011malware}. 
Beside that, it has its drawbacks:  
the main drawback of the Levenshtein algorithm is that it focuses on the global comparison between two strings, i.e., among all the variables in two strings. 
For example, the Levenshtein similarity percentage between ``John Smith'' and ``Smith, John'' equals zero. 
While we are interested in performing local similarity comparisons to find all pairs of substrings in two strings, we consider the Smith-Waterman distance as well. 

\emph{Smith-Waterman distance}: the Smith-Waterman algorithm is an alignment algorithm~\cite{newsome2005polygraph}, which gives a score to the strings. 
This score is based on a calculation of three main factor, namely match, mismatch, and the penalty score. 
A match is mostly a positive number, and it indicates that one character is aligned to a character in another string, whereas any mismatch reduces the score. 
The penalty score is reduced from the score and indicates how long one match score could continue. 
In~\cite{nauman12seq}, it is demonstrated that the maximum score is not a factor of similarity and should be further normalized. 
The authors of~\cite{arslan2001new} have addressed this by suggesting a new normalized factor based on the local distance between the position of the maximum score and the starting point, where the score is zero. 
We follow their procedure to compute the (normalized) similarity percentage between two strings. 

\newversion{\emph{Acceptable ranges of the similarity percentage}: note that in contrast to typical machine learning tasks, our approach does not aim to achieve a high similarity percentage. 
Indeed, achieving a high similarity percentage implies that the variants generated by our framework are similar to an example (e.g., a worm or a signature) previously seen by the algorithm, and consequently, can be detected by the existing NIDS. 
This is contrary to our objective that is generating unseen mutants, which are only \emph{close enough} to known examples. 
In this regard, if the similarity percentage is low (i.e., considerably less than 50\%), it means that the algorithm could not extract the pattern needed to generate a new mutant. 
}

\vspace{2pt}
\noindent\textbf{Assessing the Quality of an NIDS}

To evaluate how the quality of an NIDS in terms of attack detection is improved by employing our approach, we use the following metrics. 

\emph{False Positives (FP)}: the number of cases, where the NIDS improperly detects a flow from the benign dataset as malicious. 

\emph{False Negatives (FN)}: the number of cases, where a flow from the malicious dataset is not detected by the NIDS. 

\newversion{Furthermore, True Positives (TP) and True Negatives (TN) are defined, where the former indicates the number of cases that the NIDS correctly detects malicious flows. 
On the other hand, TN denotes the number of cases, where the NIDS correctly classifies the benign flows as ``benign.'' 
In addition to these, we use the following frequently-used metrics defined to evaluate binary classification in the context of intrusion detection\cite{szynkiewicz2017design,buczak2015survey}. 
\begin{itemize}
    \item \emph{Sensitivity}: the ratio of flows correctly classified as malicious to all malicious flows $TP/(TP + FN)$. Sensitivity is also called TP rate. 
    \item \emph{Specificity}: the ratio of flows correctly classified as benign to all benign flows $TN/(TN + FP)$.   
    \item \emph{Positive Predictive Value (PPV)}: the ratio of flows correctly classified as malicious to all items classified as malicious $TP/(TP + FP)$. 
    \item \emph{Negative Predictive Value (NPV)}: the ratio of items correctly classified as benign to all items classified as benign $TN/(TN + FN)$.
\end{itemize}

\noindent In addition to these, we report the \emph{FP rate}, i.e., the significance level defined as $1-Specificity$.  
}
\footnotesize
\begin{table}[t]
  \centering
  \begin{tabular}{|l | r|}
  \hline
    Batch size & 1  \\\hline
    Learning rate & 0.001  \\ \hline
    Number of the epochs & 100 \\ \hline
    Number of the hidden layer & 2 \\ \hline
    Word vector size & 64 \\ \hline
    Sequence Length & 1 \\ \hline
  \end{tabular}
  \caption{The LSTM configuration for our experiments.}
  \label{tab:rnn_config}
\end{table}

\footnotesize
\begin{table}[t]
  \centering
  \begin{tabular}{|l |}
  \hline
signature dpd\_ssh\_client \}\\
ip-proto == tcp\\
payload /$\hat{}$ [sS][sS][hH]-[12]\./\\
requires-reverse-signature dpd\_ssh\_server\\
enable "ssh"\\
tcp-state originator\}
\\\hline
  \end{tabular}
  \caption{Bro Signature for SSH Client protocol~\cite{bro_sig_framework}.}
  \label{tab:brosig}
\end{table}
\normalsize
\normalsize
\subsection{Setup of the Experiments}\label{sec:res_setup}
The hardware that we have used in our experiments are commercially available laptops, and a Graphics Processing Unit (GPU) server. 
The laptops act as either a user device to send traffic to the NIDS or a platform, on which Bro \newversiontwo{2.5.3} is run. 
\newversiontwo{Note that portions of Bro 2.5.3 documents, which have been used in this work, is now available in Bro 2.5.5 documents~\cite{bro2018introduction}. }
Each laptop is equipped with an Intel Core i7 - 2.6 GHz (4 Cores 8 Threads) CPU and a 16 GB DDR3 RAM, and its operating system is Linux (Ubuntu 64 Bit).

Moreover, in line with other AI-related research studies, we use GPU clusters rather than CPU (Central Processing Units) computing resources. 
This is due to the performance and computational speed of GPU clusters in comparison to CPU ones in deep learning tasks~\cite{coates2013deep,raina2009large}. 
In addition to an Intel Core i7 - 3.4 GHz (6 cores 12 Threads) CPU, our GPU server composed of two Nvidia 1080-Ti GPU cards, and a 128 GB DDR4 RAM. 
On our server, we run Linux (Ubuntu 64 Bit) and Torch accounting for deep learning tasks. 
Torch supporting Lua is selected due to the availability of many models for RNNs and LSTMs and the ability to run the algorithms on the GPUs.  
Furthermore, Torch is one of the best tools for small-scale projects and fast prototyping. 
\begin{table*}[t]
\caption{The Smith-Waterman similarity percentages between the worms generated by the XOR encoder. RNN generated mutants of three worms are labeled with a star. 
In this experiment, the worms are given to the LSTM individually and one-by-one. 
The match value equals is set to 1 (left) and 5 (right). 
The example number marked with a star shows the corresponding example generated by the LSTM.}
\label{tab:XORWorms}
\begin{minipage}{.475\linewidth}
\centering
\footnotesize
\begin{tabular}{|*{7}{c|}}
                               \cline{1-2}
  1  &100                       \\ \cline{1-3}
  2  &58.6&100                     \\ \cline{1-4}
  3  &65.5&58.4&100              \\ \cline{1-5}
  1* &50.4&48.1&49.1&100            \\ \cline{1-6}
  2* &49  &52.2&48.8&46.2&100        \\ \hline
  3* &47.8&47.9&48.6 &49.6&47.8&100 \\ \hline
     &    1 & 2 & 3 & 1* & 2* & 3* \\ \hline
\end{tabular}
\end{minipage}
\hfill
\begin{minipage}{.475\linewidth}
\centering
\footnotesize
\begin{tabular}{|*{7}{c|}}
                               \cline{1-2}
  1  &100                             \\ \cline{1-3}
  2  &60.7   &100                        \\ \cline{1-4}
  3  &66.4   &55.8    &100                 \\ \cline{1-5}
  1* &55     &49.5    &52.6  &100               \\ \cline{1-6}
  2* &51.5   &58.9    &51     &48.3    &100           \\ \hline
  3* &52     &47.2    &54.4  &53.1    &48.5   & 100 \\ \hline
     & 1  & 2      & 3    & 1*     & 2*      & 3* \\ \hline
\end{tabular}
\end{minipage}
\end{table*}
\begin{table*}[t]
\caption{Results of the experiment for Shikata Ga Nai encoder (the same setting as for the Table~\ref{tab:XORWorms}).}
\label{tab:shikataWorms}
\begin{minipage}{.475\linewidth}
\centering
\footnotesize
\begin{tabular}{|*{7}{c|}}
                               \cline{1-2}
  1  &100                             \\ \cline{1-3}
  2  &59.9 &100                        \\ \cline{1-4}
  3  &59.1 &58        &100                     \\ \cline{1-5}
  1* &49.6 &49        &51.2  &100               \\ \cline{1-6}
  2* &49.7 &49.7    &50.9  &50        &100           \\ \hline
  3* &50.6 &50.7    &54.1  &51.1    &53.4   & 100 \\ \hline
     & 1  & 2      & 3    & 1*     & 2*      & 3* \\ \hline
\end{tabular}
\end{minipage}
\hfill
\begin{minipage}{.475\linewidth}
\centering
\footnotesize
\begin{tabular}{|*{7}{c|}}
                               \cline{1-2}
  1  &100                                 \\ \cline{1-3}
  2  &58.2 &100                        \\ \cline{1-4}
  3  &59.9 &58.8    &100                         \\ \cline{1-5}
  1* &54.3 &52.6    &52.3  &100               \\ \cline{1-6}
  2* &51.6 &53.5    &50.4  &51.3    &100           \\ \hline
  3* &54.6 &51.6    &58.2  &54.2    &55   & 100 \\ \hline
     & 1  & 2      & 3    & 1*     & 2*      & 3* \\ \hline
\end{tabular}
\end{minipage}
\end{table*}
\begin{table*}[t]
\caption{Results of the experiment for the ADMMutate encoder (the same setting as for the Table~\ref{tab:XORWorms}).}
\label{tab:admWorms}
\begin{minipage}{.475\linewidth}
\centering
\footnotesize
\begin{tabular}{|*{7}{c|}}
                               \cline{1-2}
  1  &  100                           \\ \cline{1-3}
  2  &  44.8     &  100                   \\ \cline{1-4}
  3  &  47.3     &  64.2 &  100                   \\ \cline{1-5}
  1* &  52.5     &  46.6 &  48.7    &  100             \\ \cline{1-6}
  2* &  48.7     &  92.9 &  56.5    &  49.4 & 100      \\ \hline
  3* &  46.8     &  49.9 &  57        &  49.7 & 52.3  &  100  \\ \hline
    & 1    &  2  & 3& 1* &  2* & 3* \\ \hline
\end{tabular}
\end{minipage}
\hfill
\begin{minipage}{.475\linewidth}
\centering
\footnotesize
\begin{tabular}{|*{7}{c|}}
                               \cline{1-2}
  1  &  100                           \\ \cline{1-3}
  2  &  47         &  100                  \\ \cline{1-4}
  3  &  48         &  64.2    &  100               \\ \cline{1-5}
  1* &  58.2      &   47.2  &    48    &     100                 \\ \cline{1-6}
  2* &  41.2      &   96.3  &    59.8    &   38     & 100      \\ \hline
  3* &  49.6      &   55.3  &    64.1     &  51.6 & 56.8   &  100  \\ \hline
     & 1         &  2    &   3      & 1*    &  2*   &   3* \\ \hline
\end{tabular}
\end{minipage}
\end{table*}
\normalsize
\subsection{Experiment Design}\label{sec:res_design}
\vspace{2pt}
\noindent\textbf{Generating New Mutants of Polymorphic Worms}

As discussed in Section~\ref{sec:approach_worm}, we take into account the polymorphic worms generated by employing the XOR, the Shikata Ga Nai, and the ADMMutate encoders. 
Besides that, in our view, each worm is a string of bytes and depending on no protocol or server information, in line with the content-based approach taken in~\cite{li2006hamsa}. 
In this regard, we generate variants of the OS X x64 Shell Bind TCP payload~\cite{metasploit_tcp_payload}, with the different number of encoding iterations: without encoding, one iteration, etc.  
Worms mutants are then saved as a text file.  
Before feeding the text files into the LSTM, some preprocessing is needed to change the text into an understandable format for the RNN. 
Preprocessing here refers to the process of converting a text to the vectors of numbers to be understandable for the neural network. 
There are several ways to convert a text to a vector, e.g., using word to vector algorithms or decoding the text into the Unicode system. 
Selecting the appropriate model depends on the context of the data. 
For instance, using a word to vector models is more beneficial, when there exists a text written in a natural language.
Here in the preprocessing step, we use the UTF-8 encoder for the purpose of converting a text to a vector. 

\newversion{Table~\ref{tab:rnn_config} presents more details about the configuration of the LSTMs used in our experiments. 
For each input given to the LSTM (e.g., a worm), the network is configured individually based on the number of characters composing the worm. }
For instance, when a worm file contains 100 characters, the number of nodes in each hidden layer is set to 100. 
Moreover, we feed each and every worm only once. 

\vspace{2pt}
\noindent\textbf{Generating Signatures of Bro}

Bro analyzer supports several different Internet protocols. 
This means that some of the attacks launched via these protocols have been analyzed and their signatures are extracted and added into Bro signature set.  
Seventeen of these signatures related to the attacks via HTTP, SMTP, POP3, FTP, etc. are selected to feed into the LSTM. 
As an example of these signatures, Table~\ref{tab:brosig} presents the signature for SSH protocol. 
A Bro signature has mainly two parts, namely conditions and actions. 
Conditions are defined for the header and the content of the packet. 
In the content conditions, matching occurs against the payload part of the packet. 
The second part, an action, is the response given by Bro system, when a signature matching happens, e.g., rising individual events or enabling a special analyzer for the matched protocols or data~\cite{bro_sig_framework}.  

\newversion{For the experiment conducted to generate synthetic signatures, the LSTM is configured using the parameters provided in Table~\ref{tab:rnn_config}. 
The number of nodes in each hidden layer is twice the length of the signature fed into the LSTM as an input. }

\vspace{2pt}
\noindent\textbf{Post-processing the outputs of the LSTM}

The RNN generates an output according to its pre-trained model for each signature. 
These outputs have the same number of characters as the corresponding inputs. 
Each output should be recognized in Bro system as a regular expression term. 
Therefore, some editing is needed such as closing the open brackets or removing the slashes appearing in the middle of the output string. 
Theses minor modification can be easily done by an operator, or even, a scrip can be developed to perform this task. 
\newversion{In our experiments, we have followed both of these procedures: open brackets are closed by the operator, and our simple, in-house-developed script is used to remove the misplaced slashes.  }

\begin{table*}[t]
\caption{The Smith-Waterman similarity percentages between the worms generated by the XOR encoder, with the same number of iteration. 
In this experiment, we feed a set of five worms, all together, into the LSTM. 
The match value equals is set to 1 (left) and 5 (right).}
\label{tab:XORWorms_i1}
\begin{minipage}{.475\linewidth}
\centering
\footnotesize
\begin{tabular}{|*{11}{>{\centering\arraybackslash}p{0.28cm}|}}
                           \cline{1-2}
  1 & 100                       \\ \cline{1-3}
  2 & 65.3   &100                           \\ \cline{1-4}
  3 & 65.3   &62.9        & 100                 \\ \cline{1-5}
  4 & 67.6   &62.9        & 72.2   & 100            \\ \cline{1-6}
  5 & 62.9   &67.6        & 65.3   & 65.3   & 100            \\ \cline{1-7}
  1* & 48.7  & 49.8    &  49.8  &  49     & 51.7   & 100            \\ \cline{1-8}
  2* & 49.4  & 49.4    &  50.3  &  51.3  &  50.5  &  50.2 & 100            \\ \cline{1-9}
  3* & 49.9  & 50.3    &  49.7  &  49.4  &  49.8  &  50      & 49.4& 100        \\ \cline{1-10}
  4* & 47.1  & 48.1    &  47.7  &  47.7  &  48.4  &  46.2 & 49.6&  49.7  & 100    \\ \hline
  5* & 49.6  & 50.3    &  50.7  &  50.7  &  49.3  &  49.7 & 48.9&  50.1  & 51.2 & 100\\ \hline
     & 1  & 2  & 3  & 4 & 5 & 1*  & 2*  & 3*  & 4* & 5* \\ \hline
\end{tabular}
\end{minipage}
\hspace{0.5cm}
\begin{minipage}{.475\linewidth}
\centering
\footnotesize
\begin{tabular}{|*{11}{>{\centering\arraybackslash}p{0.28cm}|}}
                                                                            \cline{1-2}        
1 &100                                                                        \\ \cline{1-3}    
2 &66.5&    100                                                                     \\ \cline{1-4}
3 &66.5&    64.6 &    100                                                             \\ \cline{1-5}
4 &68.7&    64     &  72.9 &    100                                                    \\ \cline{1-6}
5 &64.8&    68.4 &    66.3 &    66.3  &    100                                            \\ \cline{1-7}
1* &54.1&    55.5 &    55.3 &    54.2    &53.4    &100                                \\ \cline{1-8}
2* &54.8&    54.5 &    55.6 &    55.6    &55        &53.9    &100                        \\ \cline{1-9}
3* &55.5&    54.5 &    56     &   56.8 &    55.5    &54.2    &55.2    &100                \\ \cline{1-10}
4* &52.7&    52.7 &    53.4 &    53.9    &52.9    &53.3    &54.3    &53.3    &100        \\ \hline
5* &54.2&    54.8 &    55.1 &    55.3    &53.9    &54.8    &54.2    &54.6    &56.1    &100\\ \hline
&1&2&3&4&5&1*&2*&3*&4*&5*                                                    \\ \hline
\end{tabular}
\end{minipage}
\end{table*}
\begin{table*}[t]
\caption{Results of the experiment for the Shikata Ga Nai encoder (the same setting as for the Table~\ref{tab:XORWorms_i1}).}
\label{tab:ShikataWorms_i1}
\begin{minipage}{.475\linewidth}
\centering
\footnotesize
\begin{tabular}{|*{11}{>{\centering\arraybackslash}p{0.28cm}|}}
                                                                            \cline{1-2}        
1&100                                                                        \\ \cline{1-3}    
2 &67.1    &100                                                                \\ \cline{1-4}
3 &61.7    &58        &100                                                        \\ \cline{1-5}
4 &58.8    &57.5    &60.8    &100                                                \\ \cline{1-6}
5 &62.6    &57.7    &59.8    &59.4    &100                                        \\ \cline{1-7}
1*&47.9    &48.1    &47.7    &47.9    &48.3    &100                                \\ \cline{1-8}
2*&48.1    &47        &47.6    &46.6    &47.5    &50.2    &100                        \\ \cline{1-9}
3*&53        &52        &52.7    &52        &52.3    &50.1    &46.7    &100            \\ \cline{1-10}
4*&48.7    &48.2    &48.5    &47.3    &47.8    &48.4    &48.9    &48.5    &100        \\ \hline
5*&47.4    &47.1    &47.1    &46.9    &46.7    &46.5    &47.1    &45.9    &49.1    &100\\ \hline
&1&2&3&4&5&1*&2*&3*&4*&5*                                                    \\ \hline
\end{tabular}
\end{minipage}
\hfill
\begin{minipage}{.475\linewidth}
\centering
\footnotesize
\begin{tabular}{|*{11}{>{\centering\arraybackslash}p{0.28cm}|}}
                                                                                \cline{1-2}        
1 &100                                                                            \\ \cline{1-3}
2 &68.2    &100                                                                    \\ \cline{1-4}
3 &62.5    &59.8    &100                                                            \\ \cline{1-5}
4 &60.2    &59.8    &61.9    &100                                                    \\ \cline{1-6}
5 &64        &59.7    &61.5    &61        &100                                        \\ \cline{1-7}
1*&53.5    &53.9    &53.7    &53.7    &53.5    &100                                    \\ \cline{1-8}
2*&52.9    &53.7    &53.8    &53.3    &56        &55.3    &100                            \\ \cline{1-9}
3*&56        &55.9    &57.3    &55.4    &54.6    &53.9    &53.6    &100                \\ \cline{1-10}
4*&55.5    &54.9    &55.1    &54.6    &54        &54.1    &54.5    &53.9    &100            \\ \hline
5*&53.8    &54.1    &52.1    &53.3    &53.4    &53.4    &52.3    &53.5    &53    &100        \\ \hline
&1&2&3&4&5&1*&2*&3*&4*&5*                                                        \\ \hline
                        
\end{tabular}
\end{minipage}
\end{table*}
\begin{table*}[t]
\caption{Results of the experiment for the ADMMutate encoder (the same setting as for the Table~\ref{tab:XORWorms_i1}).}
\label{tab:ADMWorms_i1}
\begin{minipage}{.475\linewidth}
\centering
\footnotesize
\begin{tabular}{|*{11}{>{\centering\arraybackslash}p{0.28cm}|}}
                                                                                            \cline{1-2}        
1    &100                                                                                    \\ \cline{1-3}
2    &86.1    &100                                                                            \\ \cline{1-4}
3    &90.8    &83.8    &100                                                                    \\ \cline{1-5}
4    &86.1    &83.8    &83.8    &100                                                            \\ \cline{1-6}
5    &88.5    &88.5    &83.8    &86.1    &100                                                    \\ \cline{1-7}
1*    &58        &53        &65.2    &53.4    &53.2    &100                                            \\ \cline{1-8}
2*    &52.6    &47.8    &61.2    &48        &48        &68.6    &100                                    \\ \cline{1-9}
3*    &56.6    &54.4    &65.2    &50.8    &53.1    &63.5    &65        &100                            \\ \cline{1-10}
4*    &54.4    &47.9    &63.5    &48.2    &48.1    &68.6    &87.3    &70.1    &100                    \\ \hline
5*    &54        &66.5    &52.2    &52.3    &56.3    &57.8    &55        &52.8    &55     &100               \\ \hline
&1&2&3&4&5&1*&2*&3*&4*&5*                                                                   \\ \hline
\end{tabular}
\end{minipage}
\hfill
\begin{minipage}{.475\linewidth}
\centering
\footnotesize
\begin{tabular}{|*{11}{>{\centering\arraybackslash}p{0.28cm}|}}
                                                                                            \cline{1-2}        
1    &100                                                                                    \\ \cline{1-3}
2    &86.1    &100                                                                            \\ \cline{1-4}
3    &90.8    &83.8    &100                                                                    \\ \cline{1-5}
4    &87.6    &86.6    &85.2    &100                                                            \\ \cline{1-6}
5    &89.8    &88.5    &83.8    &87.5    &100                                                    \\ \cline{1-7}
1*    &63.4    &57.9    &70.6    &58.5    &58.3    &100                                            \\ \cline{1-8}
2*    &56.8    &52        &64.4    &53.2    &52.4    &69.6    &100                                    \\ \cline{1-9}
3*    &62.1    &58.2    &70.3    &58.1    &57        &68.2    &71.1    &100                            \\ \cline{1-10}
4*    &58.3    &51.6    &67.5    &53.4    &51.9    &72.2    &90.8    &73.8    &100                    \\ \hline
5*    &58.6    &71.7    &56.6    &59.1    &60.9    &62.7    &59.4    &58.4    &59.2    &100            \\ \hline
&1&2&3&4&5&1*&2*&3*&4*&5*                                                                   \\ \hline
\end{tabular}
\end{minipage}
\end{table*}
\normalsize
\section{Results and Discussion}\label{sec:results}
\subsection{Similarity between original and RNN-generated Worms}\label{sec:res_worm_sim}
As discussed in Section~\ref{sec:res_metrics}, due to the drawbacks of the Levenshtein approach for comparing the distance between strings, we mainly focus on the results achieved by using the Smith-Waterman metric.   
Here, for the sake of completeness, we briefly present the Levenshtein similarity percentages (see~\cite{nauman12seq}) computed for our worms. 

We choose three worms generated as described in~\ref{sec:res_design} and compute the Levenshtein similarity percentages between the original worm and their corresponding RNN-generated ones. 
Note that for the XOR engine, the numbers of characters in the worms are 606, 777, and 952. 
This number is 606, 724, and 842 for Shikata Ga Nai engine and 606, 1946, and 1946 for ADMMutate engine. 
For worms encoded by the XOR engine, the average of the Levenshtein similarity percentage is 
48.9\%, whereas for the Shikata Ga Nai and ADMMutate engines, it is 49.43\% and 31.46\%, respectively. 

Table~\ref{tab:XORWorms} to Table~\ref{tab:admWorms} show the (normalized) Smith-Waterman similarity percentages between mutants of the worms themselves and the RNN generated mutants, marked with a star. 
Note that since the normalized values (over the length of the matching substrings) are reported, the length of the worms can be discarded. 
As expected, the Smith-Waterman Algorithm shows a sufficiently high similarity ratio for the worms here. 
Furthermore, when increasing the match value from one to five, the similarity percentages between worms remain in the same order of magnitude. 
Remarkable is that the similarity percentage, and accordingly the distances, between the mutants of the worms generated by the RNN (synthetic mutants) themselves, and between synthetic mutants and original ones are similar to the distances between an original mutant and other original mutants of a worm.  
This is an interesting and important result since we consider the normalized similarity percentage. 
Therefore, this result demonstrates that the LSTM not only can learn and extract the grammars underlying the given worms, but also can generate substrings with the same similarity percentage. 

It can be thought that if instead of feeding the worms one-by-one, a set of worms is given to the LSTM, the similarity percentages can be different. 
We examine this by choosing a set of 5 worms, each with a different number of encoding iterations, namely, one to five iterations. 
These worms are concatenated together into a single text file that is given to the LSTM. 
The results achieved for the XOR, Shikata Ga Nai, and ADMMutate encoders are shown in Table~\ref{tab:XORWorms_i1} to Table~\ref{tab:ADMWorms_i1}. 
In this case, an improvement in the similarity percentages (on average) can be observed. 
This can be explained by the fact that the LSTM network is definitely larger than the network used for the other experiment (Table~\ref{tab:XORWorms} to Table~\ref{tab:admWorms}) due to the larger number of characters in the text file given to that~\footnote{Recall that in our experiments the number of nodes in each hidden layer equals the number of characters in the given example.}.  
Additionally, by giving more examples to the LSTM, it can improve its prediction by observing the examples with the same grammar. 
Nonetheless, when having only one example of a worm (i.e., one mutant of that), it is still possible to obtain a close-enough unseen mutant, as presented in our previous experiments (Table~\ref{tab:XORWorms} to Table~\ref{tab:admWorms}). 
\subsection{Similarity between original and RNN-generated Bro Signatures} \label{sec:res_sig_sim}
Following the same procedure described in Section~\ref{sec:res_worm_sim}, we feed Bro signature individually into the LSTM.  
Afterwards, we compute the Levenshtein and Smith-Waterman similarity percentages between Bro signature and their associated signatures generated by the LSTM.  
In this scenario, alike the previous scenario explained in Section~\ref{sec:res_worm_sim}, the similarity percentages between the LSTM-generated signatures and the original ones demonstrate that the LSTM could find some patterns, and based on them generate a similar signature with low distance from Bro signature, see Table~\ref{tab:res_brosig}.  
It is worth noting here that the synthetic signatures should be solely \emph{sufficiently} close to the original ones. 
In other words, in order to provide an NIDS (in our case Bro) with effective signatures, we should feed new signatures that are similar to the known signatures, but also represent variants of the respective attack. 

Comparing the Smith-Waterman and the Levenshtein similarity percentages shown in Table~\ref{tab:res_brosig}, as expected, these percentages are not always consistent. 
This is due to the fact that the Levenshtein similarity percentage can reflect the results of the global comparison between two strings. 
On the contrary, the Smith-Waterman similarity percentage present the local, sub-string-based comparison between two strings.  
Although at first glance, it seems surprising that in some cases the Levenshtein similarity percentages are higher than the Smith-Waterman ones, the length of the examples (i.e., the number of the characters in the signature) can explain this. 
For shorter signatures, and smaller LSTM networks equivalently, the number of characters given to the network is not sufficient enough for the LSTM to extract and learn the grammar underlying the strings. 
And, therefore, the LSTM may repeat the same characters. 
In contrast to this for longer signatures, the LSTM extract the grammar and generate new substrings that locally match the substrings of the given signature. 
Hence, we suggest that for different signatures with various lengths, one should compute both the Smith-Waterman and the Levenshtein similarity percentages. 
\footnotesize
\begin{table}[t]  
  \centering   
  \begin{tabular}{|c |>{\centering\arraybackslash}p{1.8cm}|>{\centering\arraybackslash}p{2.2cm}|>{\centering\arraybackslash}p{2cm}|}
  \hline 
Protocol & \# Characters in signature & Levenshtein Similarity [\%] [67] & Smith-Waterman Similarity [\%] \\\hline
SOCKS & 214 & 25.23 & 81.3\\\hline
DNP3 &  38 & 84.21 & 47.6\\\hline
RFB & 30 & 80 & 35.9\\\hline
KRB & 314 & 16.24& 89.7\\\hline
FTP & 100 & 49 &  55.7\\\hline
Tunnels & 142 & 54.93 & 83.9\\\hline
DCE/RPC & 36& 75 & 29.1\\\hline
SMTP & 108 & 54.63 & 65.5\\\hline
SIP & 173 & 46.82 & 77.5\\\hline
RDP & 98 & 58.04 & 87.8\\\hline
SSH & 62 & 58.06 & 48.1\\\hline
SSL & 357 & 40.06 & 68.8\\\hline
IRC & 359 & 42.62 & 57.4\\\hline
XMPP & 72 & 80.56 & 46.3\\\hline
DHCP & 30 & 80 & 26.7\\\hline
HTTP & 406 & 78.82 & 27.7\\\hline
POP3 & 141 & 59.57 & 59.3\\\hline
  \end{tabular}
  \caption{The Levenshtein and Smith-Waterman similarity percentages between Bro signatures and the LSTM generated ones.}
  \label{tab:res_brosig}
\end{table}
\normalsize
\subsection{Enhancing the performance of Bro} \label{sec:res_bro}
Here we aim to evaluate the performance (regarding the metrics introduced in~\ref{sec:res_metrics}) of Bro, whose set of signatures is expanded by adding the synthetic signatures generated by the LSTMs. 
This section covers two series of experiment: the first set is conducted to examine how the performance of Bro can be improved in terms of detecting worms and their mutants. 
Towards the same goal, the second set of experiments is performed to observe the improvement in a general case, i.e., detecting a broad range of intrusions. 
For both of these sets, at the first stage, it is necessary to collect appropriate samples of network traffic flows. 
In our experiments, the benign data pool contains 408588 packets (330~MB in total), available in Wireshark repository~\cite{wireshark330}. 
As shown in~\cite{li2006hamsa,szynkiewicz2017design}, the size of the data pool of the benign traffic may not significantly influence the performance of an NIDS under test. 
Hence, we stick to the number of packets mentioned above. 

\begin{figure}[t]
\begin{center}
\includegraphics [width=0.8\columnwidth]{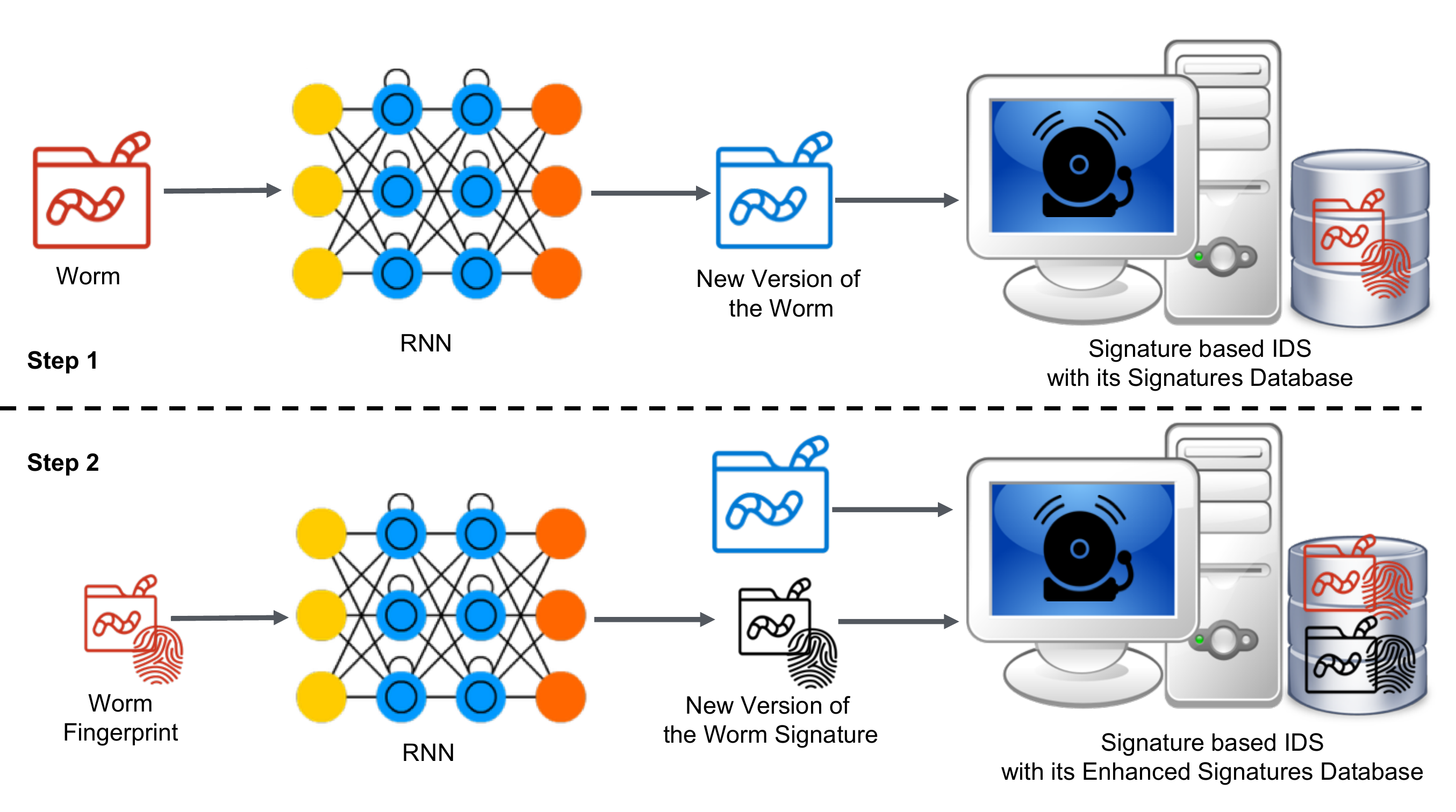}
\caption{In order to verify to what extent the performance of Bro can be improved through our approach, we first conduct Experiment~1. 
In this experiment, as the first step, we generate the mutants of a worm, which are used to evaluate the performance of Bro. 
The second step in this process is to generate a variant of the signature associated with the worm and add that to the database of Bro. }
\label{fig:experiment1}
\end{center}
\end{figure}

\noindent\textbf{Experiment~1:} As explained above, this experiment is designed to verify if integrating new, synthetic signatures into the database of Bro can improve its performance. 
For this purpose, we select the worm Code-Red that can be tracked by Bro. 
This worm exploits the buffer-overflow vulnerability to gain full system level access to its victim~\cite{moore2002code}. 
The signature associated with this worm is further available and can be used to generate a new synthetic signature for Code-Red. 
\newversion{
For this experiment, RNN is employed in two different steps and settings (see Figure~\ref{fig:experiment1}): 
\begin{itemize}
    \item Step 1: it is used to generate mutants of the worm from given examples of that. 
    This step attempts to reflect how a new mutant can be generated by an attacker. 
    Note that the results of our experiments presented in Section~\ref{sec:res_worm_sim} enable us to ensure that the mutants generated in Step~1 are sufficiently different from the original Code-Red worm. 
    \item Step 2: given the \emph{signatures} of the worm available in the Bro database, a new version of these signatures is generated by the RNN. 
    Note that this step is independent of Step~1. 
\end{itemize}
After taking these steps, it is examined if the performance of Bro can be improved in terms of detecting new worms generated in Step~1, when the database of that contains the signature generated in Step~2. }
In this regard, at the first stage, we apply Bro detector against solely the benign traffic to calculate the FP that equals 2.2\%. 
Afterwards, the worms are given to Bro, whose database is extended by adding the synthetic signature. 
As a result, in this case, all the mutants of the Code-Red are detected ($FN=0$), whereas, without the synthetic signature, $FN=16.67\%$. 
Moreover, we do not observe any change in the FP. 

\noindent\textbf{Experiment~2:} 
For this set of experiments, we generate a pool of the malicious data by using not only the signatures embedded in Bro, but also a \emph{set} of synthetic signatures. 
\newversiontwo{It is crucial to note that the set of synthetic signatures used to generate the malicious pool and the set of signatures integrated in the database of Bro are disjoint. 
In other words, we use different variant of a signature to generate the pool. }
Following a procedure commonly employed in the literature (see, e.g.,~\cite{tang2009using, szynkiewicz2017design}), both of these sets of the signatures are given to a module responsible to create malicious data from the signatures, by using, for instance, an inverse regular expression generator. 
The result is then mapped to hosts that are not involved in the benign traffic flows. 
As can be understood, we follow an overlay-like procedure to generate our pools and combine them. 
Most importantly, as advised in~\cite{aviv2011challenges}, we do not map all the malicious traffic to one host, but randomly we choose a host from a set of hosts, absent in the benign traces.
In this way, we leverage the advantages of the overlay method, namely providing a better understanding of a real-world scenario and ground truth, and accordingly a fair method to estimate the performance of an NIDS. 
Moreover, since it is possible to map more than one malicious traffic to a host, there are cases for ``concurrent infections'' caused by the same host~\cite{aviv2011challenges,stone2011underground}. 
Employing our procedure, we generate 3061 malicious traffic packet (3.2~MB in total). 
 
To evaluate the performance of Bro in terms of the FP and the FN, four different scenarios are taken into account. 
First, we apply Bro detector against solely the benign traffic to calculate the FP. 
First, similar to what we have already mentioned about Experiment~1, we test Bro in terms of FP and obtain $FP=2.2\%$.  
The second scenario is similar to the first one, but the FN of Bro is tested by giving a mixed pool of the benign and malicious traffic flows. 
In this case $FN=4.15\%$. 

In the third and fourth scenarios, we extend the set of Bro signatures by adding our LSTM-generated signatures. 
In the third case, we calculate the FP for the extended, enhanced Bro. 
The result ($FP=2.7\%$) demonstrates that adding new signatures does not considerably impair the performance of Bro in terms of the FP. 
In the fourth scenario, we calculate the FN of Bro, when our synthetic signatures are activated as well. 
In this case, our synthetic signatures are configured as Bro signatures in a signature script to match the new activities or malware. 
We obtain $FN=3.12\%$, which shows that the performance of Bro is indeed improved. 
Note that this result is achieved by adding solely a small set of new signatures, namely seventeen signatures. 
In addition to the reduction in the FN the number of the alarms raised by Bro can further illustrate the potential of our approach. 
In this regard, note that in the second scenario (without the synthetic signatures) the number of alarms is 2662, whereas in the last scenario, where the synthetic signatures are also activated, the number of alarms is 2761.   

\newversion{Furthermore, the statistical measures introduced in Section~\ref{sec:res_metrics} can be used to compare the performance of off-the-shelf (values inside the brackets) and enhanced versions of Bro as follows. 
Sensitivity is 96.65\% (95.44\%), and NPV is 96.86\% (95.89\%), where both of them show improvement as FN is reduced. 
PPV is 97.02\% (97.46\%), and Specificity is 97.21\% (97.71\%), where the degradation is related to a slight increase in FP. 
Similarly, the FP rate is increased from 2.28\% to 2.78\%. 
It is worth noting that the significantly high sensitivity and Specificity observed for the enhanced Bro make our approach a promising solution for NIDS. }

\noindent\textbf{Summary and Discussion: }
This section covers the details and results of three main experiments. 
First, we experimentally verify that the (LSTM) RNN can learn and extract the grammars underlying the given polymorphic worms, and therefore, it can be employed to generate new, unseen mutant of sophisticated polymorphic worms, with encoded exploits. 
Besides, our experimental results demonstrate that the RNN is capable of generating synthetic signatures that are sufficiently close to the original ones, and simultaneously different from the original signatures. 
These signatures represent variants of the attacks that may not be detected by an NIDS, i.e., Bro in our experiments. 
In order to show to what extent these synthetic signatures can help us to improve the performance of the NIDS, we conduct additional experiments, whose results are summarized in Table~\ref{tab:sum_res}.  
For this purpose, we provide a framework for generating a set of malicious data that involves malicious traffic flows generated from a \emph{set} of synthetic worms and signatures. 
We stress that the signatures added to the database of Bro are entirely different from the ones used to generate malicious data. 
Moreover, as the malicious traffic flows are crafted by using an inverse regular expression generator, they are further randomized. 

Finally, we put emphasis on key characteristics of RNNs, making them suitable for our purpose: RNNs can learn and extract unknown grammars and generate unseen sequences that belong to the respective grammar. 
As an example of other neural networks that may not help us with this issue, Convolutional Neural Networks (CNNs) can be mentioned. 
Although CNNs offer great potential in various applications, RNNs outperforms them in applications similar to ours, while they exhibit temporal characteristic fulfilling the conditions for processing sequences of data. 

\footnotesize
\begin{table}[t]  
  \centering   
  \begin{tabular}{|>{\centering\arraybackslash}m{4cm} |>{\centering\arraybackslash}p{2.5cm}|>{\centering\arraybackslash}p{1.2cm}|>{\centering\arraybackslash}p{1.2cm}|}
  \hline 
Experiment& NIDS & FP [\%] & FN [\%] \\\hline
\multirow{2}{4cm}{\centering Experiment~1 \hspace{2cm}(Detection of Worm Mutants)}& Off-the-shelf Bro & 2.2 & 16.67\\\cline{2-4}
& Enhanced Bro & 2.2  & 0 \\\hline
\multirow{2}{4cm}{\centering Experiment~2 \hspace{2cm}(General Case)}& Off-the-shelf Bro & 2.2 &
4.15\\\cline{2-4}
& Enhanced Bro & 2.7  & 3.12 \\\hline
  \end{tabular}
  \caption{Performance of the off-the-shelf Bro and the enhance one, evaluated in two main scenarios: detection of worms and the general case, i.e., several classes of malware.}
  \label{tab:sum_res}
\end{table}
\normalsize
\section{Conclusion and Remarks}\label{sec:conclusion}
This paper provides a framework for applying deep learning techniques in the area of cyber security. 
While zero-day attacks are vastly propagated in the network, our approach attempts to improve the ability of NIDS systems to defend against them by (1) extending their signature databases, and (2) generating a more realistic and close to the real-world ground truth to test an NIDS. 

More specifically, we take advantage of the immense power of the recurrent neural networks (RNNs) in distinguishing complex patterns in a text and generating similar ones. 
As an example of a possible application of these networks in intrusion detection, an LSTM is used to generate several mutants of polymorphic worms. 
These synthetic worms are evaluated by applying two powerful similarity metrics, namely the Levenshtein the Smith-Waterman similarity percentages. 
Furthermore, as another example of how RNNs can be beneficial to intrusion detection, we demonstrate that an LSTM can be used to generate synthetic signatures to enhance the detection rate of an NIDS. 
\newversiontwo{Nevertheless, our approach can be seen as the first step in this regard. 
More concretely, to further improve such systems, RNNIDS could benefit from methods devised to minimize false positives (see, e.g.,~\cite{hubballi2014false}), which we leave as future work. }

Last but not least, we stress that although the applications of RNNs have been widely studied and accepted in the machine learning-related literature, their capability to provide help with intrusion detection should be considered more closely. 
This paper paves the way for a more detailed exploration of this capacity in cyber security. 
\section{Acknowledgements} 
This work has been partially supported by the European Union's Horizon 2020 Research and Innovation Program under grant agreement No. 700176 (SISSDEN, Project ID: 700176, funded under: H2020-EU.3.7. - Secure societies - Protecting freedom and security of Europe and its citizens). 
Moreover, the authors would like to acknowledge the support of the Bundesministerium f{\"u}r Bildung und Forschung under grant 01\/S180251 and BIFOLD agility project. 

\bibliographystyle{ACM-Reference-Format}
\bibliography{ccs-template}

\end{document}